\documentclass[letter,12pt]{article}

\usepackage{type1cm,amsmath,amssymb,color,graphics,amscd,amsfonts,mathrsfs,epsf,indentfirst,cite}
\usepackage{epsfig}
\usepackage{bm}
\usepackage{subfigure}
\usepackage[dvipdfm,hypertex]{hyperref}
\allowdisplaybreaks[1]

\makeatletter
\@addtoreset{equation}{section}

\makeatletter
\renewcommand\section{\@startsection {section}{1}{\z@}%
                                   {-3.5ex \@plus -1ex \@minus -.2ex}
                                   {2.3ex \@plus.2ex}%
                                   {\normalfont\large\bfseries}}
\renewcommand\subsection{\@startsection{subsection}{2}{\z@}%
                                     {-3.25ex\@plus -1ex \@minus -.2ex}%
                                     {1.5ex \@plus .2ex}%
                                    {\normalfont\bfseries}}

\def\no{\nonumber \\}
\def\btab{\begin{table}[h] \begin{center} \begin{tabular}{l lp{3in}}}
      \def\etab{\end{tabular} \end{center} \end{table}}
\def\btabm{\begin{center} \begin{tabular}}
    \def\etabm{\end{tabular} \end{center}}

\def\a{{\alpha}}

\def\m{{\mu}}

\def\f#1#2{{\frac{#1}{#2}}}

\def\f {\frac}

\def\we{\wedge}

\def\CC{{\cal C}}
\def\CD{{\cal D}}

\def\CH{{\cal H}}

\def\CK{{\cal K}}

\def\CM{{\cal M}}
\def\CN{{\cal N}}

\def\CP{{\cal P}}

\def\a{\alpha}

\def\f{\phi}

\def\m{\mu}

\def\w{\omega}

\begin{document}

\begin{titlepage}
\thispagestyle{empty}
  \begin{flushright}
  PUPT-2355\\
  UT-10-19
  \end{flushright}
  
  \vspace{2cm}
		  \begin{center}
		   \font\titlerm=cmr10 scaled\magstep5
		   \font\titlei=cmmi10 scaled\magstep5
		   \font\titleis=cmmi7 scaled\magstep5
		   \centerline{\titlerm
		   Lifshitz-like Janus Solutions}

   \vspace{2cm}
   \renewcommand{\thefootnote}{$\alph{footnote}$
   }
   \noindent{{\large
   Tatsuma Nishioka$^\dagger$\footnote[1]{e-mail: nishioka@princeton.edu} and 
   Hiroaki Tanaka$^\ddagger$\footnote[2]{e-mail: tanaka@hep-th.phys.s.u-tokyo.ac.jp}
   }}\\
   \renewcommand{\thefootnote}{\arabic{footnote}}
   \vspace{1cm}

   {\it {~}$^\dagger$Department of Physics, Princeton University, Princeton, NJ 08544, USA\\
   
   \bigskip
   
   {~}$^\ddagger$Department of Physics, Faculty of Science, 
 University of Tokyo, Tokyo 113-0033, Japan
   } \\

   \vspace{1cm}
   {\large \today}
  \end{center}

  \vskip 5em

  \begin{abstract}
   We construct Lifshitz-like Janus solutions in the Einstein-scalar theory with cosmological constant
   in arbitrary dimensions.
   They are holographically dual to $z=2$ Lifshitz-like field theories with 
   a defect.
   The four-dimensional solutions can be embedded into type IIB supergravity as dilatonic deformations
   of $AdS_5 \times Y^5$ with three-form field strengths, where $Y^5$ is a five-dimensional Einstein manifold.

  \end{abstract}

\end{titlepage}

\tableofcontents

\section{Introduction}
Since the application of the AdS/CFT correspondence \cite{Maldacena:1997re,Aharony:1999ti} 
to the condensed matter physics has started, it is of great importance to find gravitational
backgrounds dual to certain non-relativistic systems.

The first example is presented in \cite{Son:2008ye,Balasubramanian:2008dm} 
with the form:
\begin{align}
	ds^2 = -r^{2z}(dx^-)^2 + 2r^2 dx^+ dx^- + r^2 d{\vec x}^2 + \frac{dr^2}{r^2} \ ,
\end{align}
where $x^+$ is compactified and $x^-$ is regarded as the time direction.
This geometry exhibits anisotropic scaling symmetry $x^- \equiv t \to \lambda^z t$, $x^+ \to \lambda^{2-z}x^+$, $x^i \to \lambda x^i$ and $r \to r/\lambda$.
The parameter $z$ is called a dynamical exponent and the symmetry of the geometry
is enhanced to the Schr{\" o}dinger symmetry including special conformal transformation
when $z=2$.
Soon after that this type of the geometry with $z=2$ was embedded into type IIB supergravity
\cite{Maldacena:2008wh,Herzog:2008wg,Adams:2008wt} deforming $AdS_5 \times S^5$ spacetime,\footnote{See also \cite{Singh:2009tq} for the embedding into the massive IIA theory.} and
the dual field theory is
identified with a discrete light cone quantization (DLCQ) of the $\CN =4$ super Yang-Mills theory.

The second example is a Lifshitz geometry of the form \cite{Kachru:2008yh}\footnote{See also \cite{Koroteev:2007yp}.}:
\begin{align}
	ds^2 = -r^{2z}dt^2 + r^2 d{\vec x}^2 + \frac{dr^2}{r^2} \ .
\end{align}
This has not the Schr{\" o}dinger symmetry but a part of it, namely, the 
non-relativistic scale invariance: $t\to\lambda^{z}t$, $x^i \to \lambda x^i$ and $r\to r/\lambda$.
This solution can be constructed in the Einstein theory coupled to 
massive gauge fields with a negative cosmological constant \cite{Kachru:2008yh,Taylor:2008tg}

Despite its simple form it is difficult to embed the four-dimensional Lifshitz solution
into massive type IIA/M-theories with warped compactification \cite{Li:2009pf,Blaback:2010pp}.\footnote{
The five-dimensional Lifshitz solution was obtained in type IIB supergravity \cite{Azeyanagi:2009pr} with the dilaton depending on $r$.
The schematic argument is given in \cite{Hartnoll:2009ns}.}
Recently the four-dimensional Lifshitz solution with $z=2$ was obtained by 
deforming the $AdS_5$ space in
null coordinates in type IIB supergravity \cite{Balasubramanian:2010uk,Donos:2010tu}:
\begin{align}\label{Lif}
	ds^2 = -2r^2 dx^+ dx^- + r^2 d{\vec x}^2 + h(x^+) (dx^+)^2 + \frac{dr^2}{r^2} \ ,
\end{align}
where we assume the five-dimensional internal manifold $Y^5$ is Einstein, {\it i.e.}, $R_{ij} = 4g_{ij}$.
When the dilaton is constant, this is just the AdS space in the null coordinates,
which is one realization of the Schr{\" o}dinger geometry \cite{Maldacena:2008wh}.
On the other hand, when the dilaton $\phi(x^+)$ depends on the null circle $x^+$ 
the function $h$ is given by $h(x^+) = (\partial_+ \phi (x^+))^2/4$ after solving the Einstein equation.
This geometry does not have the Galilean boost symmetry $x_i \to x_i -v_i x^-, ~ x^+\to x^+ 
-\frac{1}{2}(2v_i x_i - v_i^2 x^-)$ due to the $g_{++}$ component of the metric.
Thus the original Schr{\" o}dinger symmetry breaks to the Lifshitz symmetry.
After taking the Kaluza-Klein reduction along $x^+$ and identifying $x^-$ with the
time coordinate, it becomes the four-dimensional Lifshitz geometry $(Lif_4)$ with $z=2$,
although it depends on the internal coordinate $x^+$ and should be viewed as a higher-dimensional
geometry.
More realistic solutions with $h$ constant have been constructed 
by introducing the RR two-form and B-field in \cite{Donos:2010tu}, which
 can be regarded as $Lif_4 \times S^1 \times Y^5$.
Similar solutions with $z=3$ and more general solutions with $z\ge 1$ are obtained in
 \cite{Singh:2010zs} and \cite{Gregory:2010gx}, respectively.

In this paper we would like to present a holographic description of
the Lifshitz-like field theory with a defect across which the gauge coupling jumps.
Our motivation comes from a junction of two systems which 
are possible to realize in condensed matter physics.
There appear interesting phenomena in junction systems such as
Andreev reflection that is a particle scattering which occurs at an interface
between a superconductor and normal state material,
and Josephson junction that is the phenomena of electric current between 
two superconductors separated by a thin insulator.
We expect that this work will provide new insight into such systems in terms of
holographic duality in future.

The relativistic field theory with a defect can be described by the Janus solution 
that is a deformation of $AdS_5$ space with a spatially varying dilaton \cite{Bak:2003jk}.
The non-relativistic Janus with the Schr{\" o}dinger symmetry is also constructed 
by the null Melvin twist of the relativistic Janus in \cite{Karch:2009rj}.
Further developments of the Janus solution are found in \cite{Bak:2004yf,Papadimitriou:2004rz,Clark:2004sb,Clark:2005te,D'Hoker:2006uu,D'Hoker:2006uv,Hirano:2006as,Bak:2007jm,D'Hoker:2007xy,D'Hoker:2007xz,Azeyanagi:2007qj,Kim:2008dj,Gaiotto:2008sd,Honma:2008un,D'Hoker:2008wc,Chen:2008tu,Kim:2009wv,D'Hoker:2009gg,D'Hoker:2009my,Chiodaroli:2009yw,Chiodaroli:2010ur}.
We would like to construct a generalized Janus solution with a Lifshitz-like asymptotic behavior near the 
boundary.

We will start with the construction of the Lifshitz-like Janus geometry in the type IIB supergravity
by deforming AdS space with the varying dilaton, RR two-form and B-field.
Actually we find the solutions of the following form:
\begin{align}\label{LifGeo}
	ds^2 &= f(\mu)\left(d\m^2 + \frac{dz^2}{z^2} + z^2(-2dx^+dx^- + (dx_i)^2)\right)\no
& \qquad +  h(\m,x^+)(dx^+)^2 + 2g(\m,x^+)d\m dx^+ \ , 
\end{align}
where the function $g(\m, x^+)$ depends on the scalar field $\phi(\m,x^+)$ and 
the function $f(\m)$ is the same as that of the pure Janus solution. 
The function $h(\m,x^+)$ obeys a partial differential equation that cannot be 
solved analytically.
Therefore we focus on the special cases when it reduces to a ordinary differential equation.
One may worry that our solutions have an extra $g_{\m +}$ component different from the Lifshitz 
geometry, but it in fact falls off at the boundary and the metric 
approaches to \eqref{Lif}. 
To see this, we need a coordinate transformation
$(r,y)=(z\sec\m, \sin\m /z)$ ($y$ is one of the spatial direction) and use the fact
that $f(\m)=1/\cos^2\m$ when $g(\m,x^+) = 0$.
The scaling symmetry is obviously realized as $x^-\to \lambda^2 x^-,~ x_i \to \lambda x_i,~ z\to z/\lambda$.
On the other hand our solutions reduce to the Janus geometry when the dilaton 
is independent of $x^+$ and the two-forms are set to zero.
Thus our solutions have two expected aspects to describe the Lifshitz-like field theory
with a defect at $y=0$.
Although the Lifshitz solution constructed in \cite{Donos:2010tu} keeps two supersymmetries
when $Y^5$ is a Sasaki-Einstein manifold, our solution has no supersymmetry due 
to the dilatonic deformation.

The organization of this paper is as follows: we start with reviewing the constructions of
the Janus geometries of type IIB supergravity in section \ref{ss:review}. 
In section \ref{ss:LJinIIB} we construct the Lifshitz-like Janus solutions in 
type IIB supergravity.
We obtain a partial differential equation determining the function $h(\m,x^+)$.
We consider two special cases where we can solve it by an ordinary differential equation.
In section \ref{ss:LJinES} we also give an effective theory description of our solutions.
We show that our construction can be generalized to arbitrary dimensions 
with the use of the Einstein-scalar theory.
We discuss the dual field theories to our solutions in section \ref{ss:DFT}.
We conclude this paper with a discussion in section \ref{ss:discuss}.

\newpage

\section{Review of Janus solutions}\label{ss:review}
We would like to review a construction of the Janus solution 
in type IIB supergravity which is a dilatonic deformation of the $AdS_5$ spacetime 
\cite{Bak:2003jk}.
This solution does not have supersymmetry, but is still stable non-perturbatively \cite{Freedman:2003ax}.

We start with summarizing the coordinates which are suitable for the Janus geometry, and
move onto its construction.

\subsection{Coordinates}
The $(D+1)$-dimensional AdS space can be realized as a hyperboloid
\begin{align}
X_0^2 + X_{D+1}^2 - \sum_{i=1}^{D}X_i^2 = 1 \,, \label{hyperbo}
\end{align}
in the flat ${\mathbb R}^{2,D}$ space with the metric
\begin{align}
ds^2 = -dX_0^2 - dX_{D+1}^2 + \sum_{i=1}^{D}dX_i^2 \,.
\end{align}
Obviously $AdS_{D+1}$ geometry has an $SO(2,D)$ isometry.
Parametrizing $X_{D} = \tan\m$, the rest coordinates $X_{0,\cdots,D-1,D+1}$ represent the $AdS_{D}$ space with a radius $\sec\m$.
In other words, $AdS_{D+1}$ space can be sliced by $AdS_{D}$ space.
From this, the metric can be written as
\begin{align}\label{JanusCoord}
ds^2_{AdS_{D+1}} = \sec^2\m\left(d\m^2 + ds^2_{AdS_{D}} \right) \,,
\end{align}
where $\m\in [-\pi/2, \pi/2]$.
This coordinate is useful to represent the Janus geometry as we will see below. 

In the Poincar\'e coordinate, the metric is given by
\begin{align}\label{PoinCoord}
ds^2_{AdS_{D+1}} = -r^2 dt^2 + r^2\sum_{i=1}^{D-1}dx_i^2 
+ \frac{dr^2}{r^2}\ ,
\end{align}
and two coordinates \eqref{JanusCoord} and \eqref{PoinCoord} are
connected with the relations $(r=z\sec\m,\, x_{D-1}=\sin\m/z)$.
Then the boundary $r=\infty$ in the Poincar\'e coordinate is
divided into two parts $\m =\pm \pi/2$ in the coordinates \eqref{JanusCoord}.
Accordingly, the dual CFT to the AdS space can be defined on each boundary 
$x_{D-1}<0$ and $x_{D-1}>0$, respectively, in this coordinate.

\subsection{Janus solutions}
In the type IIB supergravity, the ansatz for the Janus solution takes the form,
\begin{align}
ds^2 =& f(\m)\left(d\m^2 +ds^2_{AdS_4}\right) + ds^2_{Y^5}\,, \\
\f =& \f(\m)\,, \\
F_5 =& 4\left(f(\m)^{\frac52}d\m\wedge\w_{AdS_4} + \w_{Y^5}\right)\,,
\end{align}
where $Y^5$ is a five-dimensional Einstein manifold with $R_{ij} = 4g_{ij}$.\footnote{In \cite{Bak:2003jk}, $S^5$ was chosen as the internal space but it can be generalized with such a Einstein space $Y^5$ in the same manner.}
$\w_{AdS_4}$ and $\w_{Y^5}$ are the volume forms of $AdS_4$ in unit radius and $Y^5$, respectively. 
The metric is represented in the Einstein frame.
When $f(\m)=\sec^2\m$ and $\f=$ const., it corresponds to the standard $AdS_5\times Y^5$ solution.
The dilatonic deformation breaks $SO(2,4)$ isometry of $AdS_5$ to $SO(2,3)$ of $AdS_4$.
The type IIB supergravity equations of motion are given by
\begin{align}\label{eomIIB}
&R_{MN} - \frac12\partial_M\f\partial_N\f - \frac1{96}F^2_{MN} = 0\,, \\
&\partial_M(\sqrt{-g}g^{MN}\partial_N\f) = 0\,, \\
&*F_5=F_5\,.
\end{align}
The equation of motion for the dilaton is easily solved,
\begin{align}
\partial_\m\f(\m) = \frac{c}{f(\m)^{\frac{3}{2}}}\, ,
\end{align}
with a constant of integration $c$.
The Einstein equation yields,
\begin{align}
2f'f' - 2ff'' =& -4f^3 + \frac{c^2}{2}\frac1f\,, \\
12f^2 + f'f' + 2ff'' =& 16f^3\,.
\end{align}
These two equations can be simplified to one equation,
\begin{align}\label{f'f'}
f'^2 = 4f^3 - 4f^2 + \frac{c^2}{6f}\,.
\end{align}
This equation is equivalent to the motion of a particle in the potential $V(f) = -(4f^3 - 4f^2 + c^2/6f)$ with zero energy.
To keep the factor $f$ in the metric to be positive, the top of the potential $V(f)$ must be positive.
It requires $0\leq c \leq 9/4\sqrt{2}$. 
Then \eqref{f'f'} can be integrated,
\begin{align}\label{fsol}
\m =  \pm\frac12 \int_{f_{\rm{min}}}^f d\tilde f \left(\tilde f^3 - \tilde f^2 + \frac{c^2}{24}\frac{1}{\tilde f} \right)^{-\frac12}\,, 
\end{align}
where $f_{\rm{min}}$ is the largest root of the equation $x^3-x^2+c^2/(24x)=0$.
\begin{figure}[htbp]
\centering
\includegraphics[width=5cm]{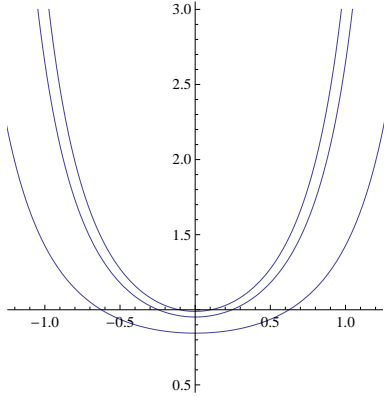}
\caption{$f(\m)$ as a function of $\m$ for $c=0.5,1,1.5$.}
\end{figure}

\begin{figure}[htbp]
\centering
\includegraphics[width=5cm]{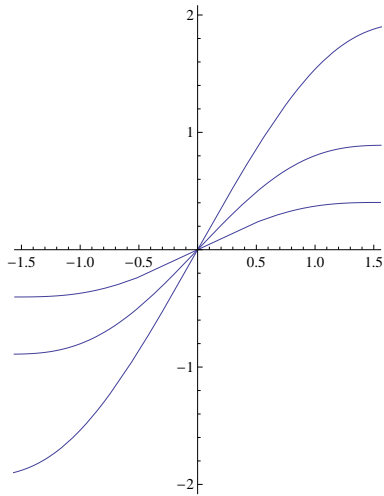}
\caption{$\f(\m)$ as a function of $\m$ for $c=0.5,1,1.5$.}
\label{Fig2}
\end{figure}

Note that taking the limit $f\to\infty$, $|\m|$ approaches to a finite constant $\m_c$.
Thus the range of $\m$ is finite: $\m \in [-\m_c,\m_c]$.
The boundary values $\m = \m_c$ and $\m = -\m_c$ correspond to two halves of the boundary of the Janus geometry.
Interestingly the dilaton approaches to two different constant values $\f(\pm \m_c)=\pm \f(\m_c)$ at the two halves as depicted in Figure \ref{Fig2}.
It makes the gauge coupling of the dual CFT jump across 
a defect inserted at $x_{3}=0$.

\section{Lifshitz-like Janus solutions in type IIB supergravity}\label{ss:LJinIIB}

Now we describe Lifshitz-like generalization of the Janus solution 
in type IIB supergravity. 
Firstly, we start with general cases under the assumption of the
appropriate metric form. Then we obtain a partial differential equation
from the Einstein equation,
but it cannot be solved analytically. Therefore, we focus on two cases
where it reduces to an ordinary differential equation.
Our solutions will become the Lifshitz geometries constructed in
\cite{Balasubramanian:2010uk} for the case 1
and a part of those given in \cite{Donos:2010tu} for the case 2, respectively.

\subsection{Generalities}
We set the axion zero for simplicity.
In the Einstein frame, the action
including a dilaton, B-field, RR three-form and RR five-form is
\begin{align}
	I = \frac{1}{2\kappa^2_{10}} &\int d^{10}x \sqrt{-g} \left(  R - \frac{1}{2}(\partial \phi)^2
	- \frac{1}{2}e^{-\phi}|H_3|^2 - \frac{1}{2}e^{\phi}|F_3|^2  - \frac{1}{4}|\tilde F_5|^2\right)\no
	& \qquad - \frac{1}{4\kappa^2_{10}} \int C_4\we H_3\we F_3  \ ,
\end{align}
where $\tilde F_5 = dC_4 - \frac{1}{2}C_2 \we H_3 + \frac{1}{2}B_2 \we F_3$.
The equations of motion are 
\begin{align}
	& \Box \phi = -\frac{1}{12}(H_3)^2 e^{-\phi} + \frac{1}{12}(F_3)^2 e^{\phi} \ , \\
	&R_{MN} - \frac{1}{2}\partial_M\phi\partial_N\phi - \frac{1}{12}e^{-\phi}\left(3(H_3^2)_{MN} - \frac{1}{4}(H_3)^2 g_{MN} \right) \no
    &\qquad -\frac{1}{12}e^{\phi}\left(3(F_3^2)_{MN} - \frac{1}{4}(F_3)^2 g_{MN} \right) - \frac{1}{96}(\tilde F_5^2)_{MN} = 0 \ .
\end{align}
The fluxes satisfy the Bianchi identities:
\begin{align}
dH_3 =& dF_3 = d\tilde F_5 - H_3 \we F_3 = 0 \ , 
\end{align}
and the equations of motion:
\begin{align}
d (e^{-\phi}\ast H_3) = 0 \ , \qquad d(e^{\phi}\ast F_3) = 0 \ , \qquad d\ast  \tilde F_5 = H_3 \we F_3  \ ,
\end{align}
and the self-dual condition for $\tilde F_5$
\begin{align}
\tilde F_5 =& \ast \tilde F_5 \ .
\end{align}

We assume solutions with the following form:
\begin{align}\label{Ansatz}
	ds^2 =& f(\mu)\left(d\m^2 + \frac{dz^2}{z^2} + z^2(-2dx^+dx^- + dx_1^2)\right) \no
 & \qquad + h(\m , x^+)(dx^+)^2  + 2g(\m, x^+)dx^+ d\m + ds^2_{Y^5}\,, \\
\f =& \f(\m,x^+)\ , \\
H_3 =& dx^+ \we W_1 \ , \\
F_3 =& dx^+ \we W_2 \ , \\
\tilde F_5 =& 4\left(\w_{5} + \w_{Y^5}\right)\ ,
\end{align}
where $Y^5$ is a five-dimensional Einstein manifold with $R_{ij} = 4g_{ij}$ and $W_a$ $(a=1,2)$ are two-forms
defined on $Y^5$.

Given this ansatz, $(H_3)^2$ and $(F_3)^2$ vanish and the two-forms $W_a$ 
satisfy:
\begin{align}
	&dx^+\we dW_a = d\ast_{Y^5} W_a = 0 \ .
\end{align}
These conditions are satisfied when $W_a$ are harmonic two-forms on $Y^5$.

The equation of motion for the dilaton is easily solved,
\begin{align}
\partial_\m\f(\m, x^+) = \frac{c(x^+)}{f(\m)^{\frac32}}\, ,
\end{align}
with introducing a new function $c(x^+)$.
The $\m\m$, $+-$, $zz$ and $yy$ components of the Einstein equation give the equations that resemble the Janus case.
\begin{align}
2f'f' - 2ff'' =& -4f^3 + \frac{c^2(x^+)}2\frac1f\,, \label{f1}\\
12f^2 + f'f' + 2ff'' =& 16f^3\,. \label{f2}
\end{align}
It follows from the first equation that the function $c(x^+)$ must be constant: $c(x^+) = c$.
Thus $f(\m)$ is exactly given by \eqref{fsol} and $\phi$ can be represented as
\begin{align}
\phi (\m,x^+) = c \int \frac{d\m}{f(\m)^{\frac{3}{2}}} + \phi^{(+)}(x^+) \ .
\end{align}
The range of $\m$ and the possible value of $c$ are the same as the Janus case.

\begin{figure}[htbp]
\centering
\includegraphics[width=6cm]{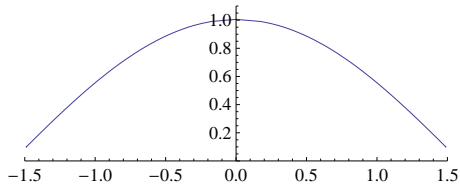}
\caption{
$f(\m)^{-\frac{1}{2}}$ as a function of $\m$ for $c=0.5$, to which $g(\m)$ is proportional.}
\label{Fig3}
\end{figure}

The $+\m$ component of Einstein equation determines the function $g$ as
\begin{align}\label{g}
g(\m,x^+) =\frac{c}{6}  \frac{ \partial_+ \phi(x^+)}{ f(\m)^{\frac12}}\,.
\end{align}
Then $g(\m,x^+)$ becomes zero at the boundary $\m=\pm\m_c$
as shown in Figure \ref{Fig3}.

The $+z$ component is automatically satisfied from this solution.

The $++$ component will be a differential equation for $h$:
\begin{align}\label{DEH}
	&\frac{\partial_\m^2 h(\m,x^+)}{2f} - \frac{f'}{4f^2}\partial_\m h(\m,x^+) - \left( 2 - \frac{c^2}{12 f^4}  \right) h(\m,x^+) \no
	& \qquad + \left( \frac{1}{2} - \frac{c^2}{18 f^3}\right)(\partial_+\f^{(+)})^2  +\frac{(W_1)_{ij}(W_1)^{ij}}{4} e^{-\phi(\m,x^+)} +
	\frac{(W_2)_{ij}(W_2)^{ij}}{4} e^{\phi(\m,x^+)} = 0 \ .
\end{align}
Given the parameter $c$ and the harmonic two-forms $W_\a$, we can solve this partial differential equation and obtain
the function $h(\m,x^+)$ numerically.
The term that involves $dx^+$ in the metric \eqref{Ansatz} can be written as
\begin{align}
h(\m,x^+)\left(dx^+ - z^2\frac{f(\m)}{h(\m,x^+)}dx^- + \frac{g(\m,x^+)}{h(\m,x^+)}d\m\right)^2 - \frac{1}{h(\m,x^+)}(z^2f(\m)dx^- - g(\m,x^+)d\m)^2 \ .
\end{align}
Compactification of the first bracket direction yields the Lifshitz-like geometry with 
the dynamical exponent $z=2$ near the boundary.
More precisely, our solution  
has a scaling symmetry $t\equiv x^- \to \lambda^2 x^-, ~
x^1\to \lambda x^1, ~z\to z/\lambda$ with $x^+$ fixed, 
where $x^-$ is identified with a new time coordinate.
We, however, will focus on the special cases such that the analytic solutions can be constructed below.

\subsection{Case 1: $W_a=0$}\label{ss:LJBN}
\begin{figure}[htbp]
\centering
\includegraphics[width=5cm]{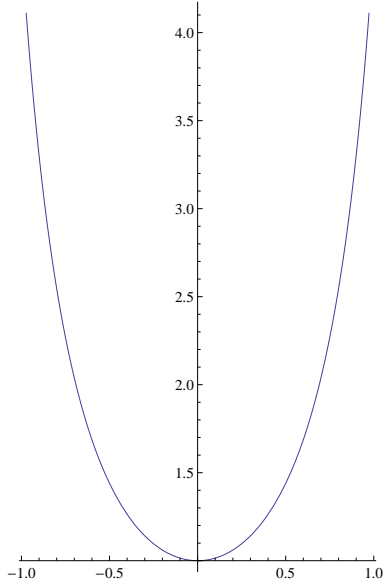}
\caption{$\tilde h(\m)$ is proportional to $f(\m)$ at the boundary as a function of $\m$ for $c=0.5$.}
\label{Fig4}
\end{figure}

\begin{figure}[htbp]
\centering
\includegraphics[width=5cm]{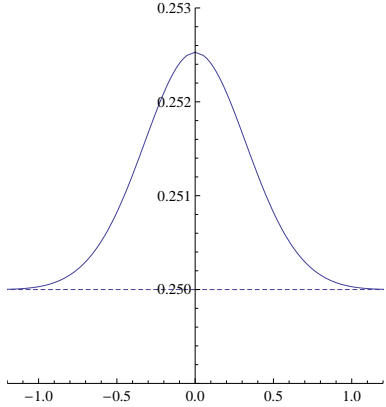}
\caption{
$\tilde h(\m)$ approaches to $1/4$ at the boundary as a function of $\m$ for $c=0.5$.}
\label{Fig5}
\end{figure}

We can simplify the partial differential equation \eqref{DEH} when $W_a =0$, and
reduce it to the ordinary differential equation further requiring $h(\m,x^+) = \tilde h (\m) (\partial_+
\phi^{(+)})^2$:
\begin{align}\label{tildeh}
&\frac{\tilde h''}{2f} - \frac{f'}{4f^2} \tilde h' - \left( 2 - \frac{c^2}{12 f^4}  \right) \tilde h + \left( \frac{1}{2} - \frac{c^2}{18 f^3}\right)= 0  \ .
\end{align}
Expanding $\tilde h$ in $1/f$ as
\begin{align}
\tilde h(\mu) = \sum_{k=k_0}^{\infty}\frac{a_k}{f(\mu)^{k}} \ ,
\end{align}
and using \eqref{f1}, \eqref{f2}, and \eqref{tildeh}, we can obtain an equation without the differentials of $f$,
\begin{align}
\left( \frac{1}{2} - \frac{c^2}{18 f^3}\right)
+ 2a_{k_0}(k_0^2-1)f^{-k_0} & \no
+ \left(-a_{k_0}k_0(2k_0 + 1) + 2a_{k_0+1}((k_0+1)^2 - 1)\right)f^{-(k_0+1)} & \no
+ \left(-a_{k_0+1}(k_0+1)(2(k_0+1) + 1) + 2a_{k_0+2}((k_0+2)^2 - 1)\right)f^{-(k_0+2)} & \no
+ \left(-a_{k_0+2}(k_0+2)(2(k_0+2) + 1) + 2a_{k_0+3}((k_0+3)^2 - 1)\right)f^{-(k_0+3)} & \no
+ \sum_{k=k_0}^\infty \left(\frac{c^2a_k}{12}(k+1)^2 - a_{k+3}(k+3)(2(k+3) + 1) + 2a_{k+4}((k+4)^2 - 1)\right)f^{-(k+4)} & = 0 \ .
\end{align}
It determines the power of the leading term $k_0=0,-1$ and the coefficients $a_i$ are determined recursively.
Thus there are two types of solutions near the boundary of $\m$
where $f(\m)$ diverges. One is proportional to $f$, {\it i.e.}, $k_0=-1$ (Figure \ref{Fig4}):
\begin{align}
	\tilde h(\m) \simeq \a f(\m) \qquad (\m \to \pm \m_c) \ ,
\end{align}
and the other approaches constant, {\it i.e.}, $k_0=0$, $a_{k_0}=1/4$ (Figure \ref{Fig5}):
\begin{align}
	\tilde h(\m) \simeq \frac{1}{4} \qquad (\m \to \pm \m_c) \ .
\end{align}
In the latter case, the function $h(\m,x^+)$ approaches to $(\partial_+\phi^{(+)}(x^+))^2/4$, and the function $g(\m,x^+)$ given by \eqref{g} is going to zero near the boundary. 
Then our solution actually behaves like the Lifshitz solution given by \cite{Balasubramanian:2008dm}, while the dilaton takes different values on each half of the boundary similarly to the Janus solution.
Therefore we realize the Lifshitz-like Janus solution in our setup.

\subsection{Case 2: $\partial_+\phi^{(+)}=0$}\label{ss:LJGD}
The second example is a solution with $\partial_+\phi^{(+)}=0$, when $g(\m,x^+)=0$ and
the partial differential equation \eqref{DEH} reduces to an ordinary differential equation
under the assumption such that $h$ and $W_a$ do not depend on $x^+$.

Thus we can solve the equation once the parameter $c$ is given.
When $c=0$, the dilaton becomes constant and \eqref{DEH} has a simple solution $h=(W_1)_{ij}(W_1)^{ij}e^{-\phi}/8+(W_2)_{ij}(W_2)^{ij}e^\phi/8$, which
is included in the solutions given in \cite{Donos:2010tu}.
When $c\neq 0$, we have a solution with following behaviors around the boundary:
\begin{align}
	&h(\m\to \pm \m_c) = \frac{(W_1)_{ij}(W_1)^{ij}}{8} e^{-\phi (\pm\m_c)} + \frac{(W_2)_{ij}(W_2)^{ij}}{8} e^{\phi (\pm\m_c)} \ .
\end{align}
Therefore our solution is the Lifshitz-like Janus geometry with a constant $h$ near 
the boundary.
If we take $W_1 = W_2$, the value of $h$ at each half of the boundary is equal  
because the dilaton $\phi(\m)$ is the odd function of $\m$.

As opposed to the previous case, $Y^5$ cannot be arbitrary but
should be a manifold which has a non-trivial two-cycle to define the harmonic two-forms
on it.
The concrete example is the Sasaki-Einstein space $T^{1,1}$ \cite{Romans:1984an,Candelas:1989js} 
with the metric of the form:
\begin{align}
	ds^2_{T^{1,1}} = \frac{1}{6} \sum_{i=1}^2(d\theta_i^2 + \sin^2\theta_i d\phi_i^2)
	+\frac{1}{9}(d\psi + \cos\theta_1 d\phi_1 + \cos\theta_2 d\phi_2)^2 \ ,
\end{align} 
where $\psi \in [0, 4\pi )$.
It is obvious that $T^{1,1}$ is a $S^1$ fibration on $S^2 \times S^2$.
We can choose the harmonic two-forms on $T^{1,1}$ as
\begin{align}
	W_1 = W_2 = \frac{w}{2} (\sin\theta_1 d\theta_1 \we d\phi_1 - \sin\theta_2 d\theta_2 \we d\phi_2) \ ,
\end{align}
where $w$ is a constant we can tune freely.
Then the value of $h$ at the boundary is
\begin{align}
	h(\m = \pm\m_c) = 9 w^2 \cosh \phi(\m_c) \ ,
\end{align}
and the radii of the circle $x^+$ are equal on each boundary.

\section{Effective theory for Lifshitz-like Janus solutions}\label{ss:LJinES}
We have constructed the Lifshitz-like Janus solutions in type IIB supergravity so far.
It is worthwhile to consider generalization of our solutions to arbitrary dimensions
in effective theories, although we don't know how they can be embedded 
into the string/M-theories.

We consider the Einstein-Hilbert action with one neutral and one complex scalars
 assuming a similar form of the metric. The role of the B- and RR-fields is played by the 
 complex scalar field
which gives the energy-momentum tensor with only $++$ component.\footnote{We are grateful to C. Herzog for 
useful discussions of this construction.}

\subsection{Einstein-scalar theory}

The key to obtain the Lifshitz-like solution is introducing the 
energy-momentum tensor with only $++$ component.
Let us consider Lifshitz-like Janus solutions in the $(D+1)$-dimensional 
Einstein-Hilbert theory with one neutral and one complex scalars
\begin{align}\label{action}
	I = \int d^{D+1}x \sqrt{-g} \left( R + \Lambda - \frac{1}{2}(\partial\phi)^2 - \frac{1}{2}|\partial\chi |^2  \right) \ ,
\end{align}
When $\Lambda = D(D-1)/\ell^2$ there exists an $AdS_{D+1}$ solution with a curvature radius $\ell$ with constant scalar fields. 

The equations of motion are the followings:
\begin{align}
	&R_{MN} = -D g_{MN} + \frac{1}{2}\partial_M\phi\partial_N\phi + \frac{1}{4}(\partial_M\chi\partial_N\bar\chi + \partial_M\bar\chi\partial_N\chi) \ , \label{eomEin}\\
	&\Box\phi = \Box\chi
	= 0 \ .\label{eomsca}
\end{align}

The Lifshitz-like Janus solution 
we consider is a deformation of the $AdS_{D+1}$ space with varying dilaton
and we take the following ansatz setting $\ell = 1$:
\begin{align}
	ds^2 &= f(\mu)\left(d\m^2 + \frac{dz^2}{z^2} + z^2(-2dx^+dx^- + (dx_i)^2)\right)\\  &\qquad\qquad + h(\m,x^+)(dx^+)^2 + 2g(\m,x^+)d\m dx^+  \ , \label{LJD}\no
\f =& \f(\m,x^+)\,, \\
\chi =& \chi(x^+) \ , \label{anschi}
\end{align}
where $i$ runs from $1$ to $D-3$.
The equation of motion for the complex scalar is automatically satisfied.

This situation is very similar to the previous case once $W$ is replaced with $\chi$.
The complex scalar field contributes to only the $++$ component of the energy-momentum tensor.
Then we obtain Lifshitz-like Janus solutions similarly to the 
type IIB case.

The components of the Ricci tensor we will use below are as follows
\begin{align}
	R_{\m\m} &= \frac{D (f'^2 - f f'')}{2 f^2} \ , \\
	R_{zz} &= -\frac{2ff'' + (D-3)f'^2 + 4(D-1)f^2}{4z^2 f^2} \ , \\
	R_{\m +} &= -\frac{g ((D-3)f'^2 + 2ff'')}{4f^3} \ ,\\
	R_{z+} &= -\frac{2f\partial_\m g + (D-3)g\partial_\m f}{2zf^2} \ , \\
	R_{++} &= -\frac{1}{4f^3} \left(2f^2 \partial_\m^2 h - (5-D)ff'\partial_\m h + 2(f'^2 + 4f^2)h -f(8g^2 + 2(D-3)f'\partial_\m g + 4f\partial_\m^2 g)\right) \ , \\
	R_{+-} &= -z^4R_{zz} \ , \qquad R_{ii} = z^4 R_{zz} \ .
\end{align}

The ansatz \eqref{anschi} automatically solves the equation \eqref{eomsca} for the complex scalar.
The equation \eqref{eomsca} for the real scalar is solved 
\begin{align}\label{scsol}
	 \phi (\m , x^+) = c(x^+)\int \frac{d\m}{f(\m)^{\frac{D-1}{2}}}  + \phi^{(+)}(x^+)\ .
\end{align}
The $\m\m$ part and the $+-$, $zz$ and $ii$ components of the Einstein equation give the equations 
\begin{align}
D(f'^2 - ff'') =& -2D\, f^3 + \frac{c(x^+)^2}{f^{D-3}}\,, \\
4(D-1)f^2 + (D-3)f'^2  + 2ff'' =& 4D \,f^3\,. \label{zz}
\end{align}
It follows from the first equation that $c(x^+)$ must be constant and thus 
the function $f(\m)$ is obtained by the following relation
\begin{align}
	&c(x^+) = c = \mathrm{const} \ , \\
	&\m = \pm \frac{1}{2} \int_{f_{\rm min}}^{f} d\tilde f \left( \tilde f^3 - \tilde f^2 + \frac{c^2}{2D(D-1)\tilde f^{D-3}} \right)^{-\frac{1}{2}} \ ,
\end{align}
where $f_{\rm min}$ is given by the largest root of the equation $x^3-x^2 +c^2/(2D(D-1)x^{D-3})=0$.
In the same manner as the last section, the parametric range of $c$ is 
\begin{align}
0 \leq c \leq \sqrt{\frac{2(D-1)^D}{D^{D-1}}} \ .
\end{align}
The $\m +$ part of the Einstein equation is
\begin{align}
	-\frac{g}{4f^3}((D-3)f'^2 + 2ff'') = -D g + \frac{1}{2}\partial_\m \phi \partial_+\phi \ ,
\end{align}
and it can be simplified by using \eqref{scsol} and \eqref{zz} to
\begin{align}
	g(\m, x^+) = \frac{c \,\partial_+ \phi^{(+)}}{2(D-1) f^{\frac{D-3}{2}}} \ .
\end{align}
The $z +$ component is automatically satisfied by the above solution.

Finally we have to solve the $+ +$ component but it is a partial differential equation. 
\begin{align}\label{ETh}
\frac{1}{2f} \partial_\m^2 h &- \frac{(5-D)}{4f^2} f' \partial_\m h + \left( (2-D) + \frac{c^2}{D(D-1)f^{D}}\right) h \no
&+ \left( \frac{1}{2} - \frac{c^2}{2(D-1)^2f^{D-1}} \right) (\partial_+\phi^{(+)})^2 + \frac{1}{2} |\partial_+\chi|^2 = 0 \  .
\end{align}

Solving this differential equation, we obtain Lifshitz-like Janus solutions in arbitrary dimensions. Corresponding to the special cases in type IIB supergravity, 
it can be reduced to an ordinary differential equation 
when appropriate fields are set to zero.
It is worth mentioning that the function $g(\m, x^+)$ goes to zero near the
boundary when $D\ge 4$, and then our solution actually behaves as a $(D+1)$-dimensional 
Lifshitz solution.

In the case of $\partial_+ \chi = 0$ the above equation is simplified to an ordinary differential 
equation with $h(\m , x^+) = \tilde h(\m) (\partial_+\phi^{(+)})^2 $.
When $c=0$ this solution becomes a Lifshitz-like geometry with $f(\m) = \sec^2 \m$ and
\begin{align}
	\tilde h(\m) = \frac{1}{2(D-2)} \ .
\end{align}
When $c\neq 0$ we still have a solution which approaches 
the above value near the boundary $\m = \pm\m_c$.

On the other hand, we have another class of solutions when $\partial_+ \phi^{(+)}=0$
corresponding to the case 2 considered in section \ref{ss:LJGD}.
We need to take $|\partial_+ \chi|^2$ constant to reduce \eqref{ETh} to an ordinary 
differential equation of $h$ with respect to $\m$. 
Then the boundary value of $h$ is 
\begin{align}
	h(\m=\pm \m_c) = \frac{|\partial_+ \chi|^2}{2(D-2)} \ .
\end{align}
Notice that $g$ is zero in this case and $\chi$ must be complex to take 
$|\partial_+ \chi|^2$ constant since $x^+$ is compactified to $S^1$.

\section{Dual field theories}\label{ss:DFT}
The $AdS_5 \times S^5$ background with a constant dilaton in type IIB supergravity
is dual to the $\CN =4$ super Yang-Mills theory in four dimensions
 with an appropriate gauge coupling.
Once we deform the background with varying the dilaton, the gauge coupling in 
the dual field theory depends on the spacetime coordinates. 
More precisely the relation between the gauge coupling and the dilaton is 
given by $g_{YM}^2(x^M) = e^{\phi (x^M)}$ where $x^M$ denotes the spacetime coordinates at the boundary. 

The Lifshitz-like solution with $\phi = \phi(x^+)$ is 
a light-like deformation of the $AdS_5$ with the nontrivial dilaton $\phi(x^+)$.
This solution exhibits anisotropic scaling symmetry with the dynamical exponent $z=2$ such that $x^-\to \lambda^2 x^-$, $x_i \to \lambda x_i$ and $z \to \lambda^{-1}z$. In addition to this dilatation symmetry $\CD$, there are time translation $\CH$,
spatial translations ${\CP}^i$ and spatial rotations $\CM^{ij}$.
This symmetry algebra closes on itself. 
If the dilaton is constant, the background enjoys the 
Schr{\" o}dinger symmetry including the Galilean boosts $\CK^i$, special conformal transformation $\CC$ and mass or particle number operator $\CM$.
Now we don't have these extended symmetries due to the $g_{++}$ component in the metric.

The Janus solution with $\phi = \phi(\m)$ has the $SO(3,2)$ symmetry which is the subgroup of the $SO(4,2)$ symmetry of $AdS_5$.
The dilaton varies in the bulk and takes different values on each boundary $\m = \pm\m_c$. 
This means that the dual field theory is the $\CN =4$ SYM living on the boundary consisting of two half
spaces $\m = \pm \m_c$, where the coupling constants are given by
$g_{YM}^2(\pm\m_c) = e^{\phi(\pm\m_c)}$.
In the Poincar\'e coordinates $(r=z\sec \m, x_3=\sin\m /z)$, the two halves of the boundary 
are glued at $x_3 = 0$ and the gauge coupling jumps across a defect at $x_3=0$.
Although the four-dimensional conformal symmetry is broken, there still
exists three-dimensional conformal symmetry $SO(3,2)$ transverse to the defect.
In addition the dilatation symmetry enhances and acts as $t \to \lambda t$, and $x_i \to \lambda x_i$ as
well as $x_3 \to \lambda x_3$.
This symmetry algebra doesn't change the location of the defect and 
thus the dual field theory is called a defect CFT.

The $(D+1)$-dimensional Lifshitz-like Janus solution in the effective theory interpolates 
these two types of solutions and dual to the $(D-1)$-dimensional Lifshitz-like defect CFT 
with the dynamical exponent $z=2$.
In the type IIB context, the dual field theory for the case 1 is the DLCQ of $\CN =4$ SYM 
with a defect since $x^+$ coordinate is compactified, while
we don't understand the specific structure of the dual field theory for the case 2.
We will discuss it in section \ref{ss:discuss}.

Before concluding this section, we would like to mention the difference between our
solution and the non-relativistic Janus solution constructed in \cite{Karch:2009rj}.
The situation is very similar, but the latter is dual to the defect CFT with the Schr{\" o}dinger 
symmetry that is larger than that of our solutions.

\section{Discussion}\label{ss:discuss}

It seems to be possible to embed three-dimensional Lifshitz-like Janus solutions into 
M-theory since 
the action \eqref{action} we used is fairly simple.
But it is actually hard to obtain a massless scalar field by compactifying IIA/M-theory on a six-dimensional internal manifold.
Fluxes in type II theory prevents the dilaton from being massless in general although a self-dual flux doesn't.
This is why we can find the Lifshitz-like Janus solution in type IIB case.

One may think such a solution can be obtained by deforming the $AdS_4 \times Y^7$ spacetime
with the four-form flux
\begin{align}
	F_4 = 3\omega_4 + d\phi (x^+,\m) \we W \ ,
\end{align}
where $W$ is a three-form defined on the internal manifold $Y^7$.
Unfortunately, this way of construction doesn't work as described below.
The equations of motion and the Bianchi identity for the flux 
are satisfied if $\Box \phi = 0$, $dW=0$ and $d\ast_7 W=0$.
The former is the equation of motion for a massless scalar and the latter
means that $W$ is a harmonic three-form on $Y^7$.
On the other hand the internal component of the Einstein equation
requires $(\partial\phi)^2 = 0$, but this cannot be hold on our background
\eqref{LJD} unless $\phi$ is independent of $\m$, which finally reduces to the
Lifshitz geometry given in \cite{Donos:2010tu} and the appendix A.1 of \cite{Balasubramanian:2010uk}.

Our solutions are non-supersymmetric due to the dilatonic deformation.
Then they could be unstable against perturbation. 
There exists supersymmetric Lifshitz \cite{Donos:2010tu} and Janus solutions 
\cite{Clark:2005te,D'Hoker:2006uu,D'Hoker:2007xy,D'Hoker:2007xz,D'Hoker:2008wc,D'Hoker:2009gg,D'Hoker:2009my,Chiodaroli:2009yw}, 
respectively.
Thus it must be interesting to investigate the construction of supersymmetric 
Lifshitz-like Janus solutions in string/M-theories.
Our effective theory construction might be useful for this purpose.

In the case 2 with $Y^5 = T^{1,1}$ in section \ref{ss:LJGD}, our solution becomes a deformation of Klebanov-Witten theory
\cite{Klebanov:1998hh} with the dilaton, the B- and RR 2-form fluxes on $T^{1,1}$. 
When the dilaton is constant, our solution reduces to that of \cite{Donos:2010tu}, which
consists of D3-branes on a conifold and $(p,q)$ five-branes wrapped over a three-cycle
in $T^{1,1}$.
The remaining three dimensions of the $(p,q)$ five-branes parallel to ${\mathbb R}^{1,3}$, and then
it would represent a light-like domain wall in the dual field theory similarly to \cite{Klebanov:2000hb,Herzog:2001xk}.
The Lifshitz-like field theory is expected to be realized on this domain wall, although 
we have no evidence for it except the scaling symmetry of our bulk geometry.
It will be interesting to investigate this further.

More interesting things can happen if there is a black hole inside our solutions.
Dual field theories will be heated up and there would be phase transitions
in the presence of matters.
Unfortunately, black hole solutions have not been constructed even in the Janus
geometry except the special case \cite{Bak:2007jm}. 
It is of great importance to construct such solutions in order to holographically describe 
more realistic field theories related with condensed matter physics.

 \vspace{1.3cm}
 \centerline{\bf Acknowledgements}
 We are grateful to C. Herzog, T. Ishii, I. Klebanov, Y. Tachikawa and M. Yamazaki for valuable discussions.
 TN would like to thank all members of the High Energy Physics Theory Group of 
 the University of Tokyo for the hospitality, and the Galileo Galilei Institute for Theoretical 
 Physics for the hospitality and the INFN for partial support 
 during the completion of this work.
The work of TN was supported in part by the US NSF under Grants No.\,PHY-0844827 and
PHY-0756966.
The work of HT was supported by Global COE Program ``the Physical Sciences Frontier", MEXT, Japan.



\end{document}